\documentclass[preprint]{aastex}
\usepackage{rotating}
\usepackage{natbib}
\slugcomment{Accepted by ApJ: October 24th, 2013.} 

\def\apjs{{\rm ApJS}}                  
\def\apjl{{\rm ApJ Let}}               

\def\a4{\hsize 17.0cm \vsize 25.cm}

\shorttitle{Cooling by dust in young superstar clusters}
\shortauthors{Tenorio-Tagle et al.}
\received{July 30th, 2013}
\accepted{October 24th, 2013}

\begin{document}

\title{Dusty supernovae running the thermodynamics of the matter reinserted within young and massive super stellar clusters}

\author{Guillermo Tenorio-Tagle\altaffilmark{1}, Sergiy Silich\altaffilmark{1}, 
Sergio Mart\'inez-Gonz\'alez\altaffilmark{1}, Casiana Mu\~noz-Tu\~n\'on\altaffilmark{2,3},  
Jan Palou\v{s}\altaffilmark{4} and Richard 
W\"unsch\altaffilmark{4}}

\altaffiltext{1}{Instituto Nacional de Astrof\'\i sica \'Optica y
Electr\'onica, AP 51, 72000 Puebla, M\'exico; gtt@inaoep.mx}
\altaffiltext{2}{Instituto de Astrof\'{\i}sica de Canarias, E 38200 La
Laguna, Tenerife, Spain; cmt@ll.iac.es}
\altaffiltext{3}{Departamento de Astrof\'\i sica, Universidad de La Laguna, 
E-38205, La Laguna, Tenerife, Spain}
\altaffiltext{4}{Astronomical Institute, Academy of Sciences of the Czech
Republic, Bo\v{c}n\'\i\ II 1401, 141 31 Prague, Czech Republic}

\begin{abstract}
Following the observational and theoretical evidence that points at core 
collapse supernovae as major producers of dust, here we calculate the 
hydrodynamics of the matter reinserted within young and massive super stellar 
clusters under the assumption of gas and dust radiative cooling.  The large 
supernova rate expected in massive clusters allows for a continuous 
replenishment of dust immersed in the high temperature thermalized 
reinserted matter and warrants a stationary presence of dust within the 
cluster volume during the type II supernova era. We first show that such a 
balance determines the range of dust to gas mass ratio and this the dust 
cooling law. We then search for the critical line that separates stationary 
cluster winds from the bimodal cases in the cluster mechanical luminosity 
(or cluster mass) {\it vs} cluster size parameter space. In the latter, 
strong radiative cooling reduces considerably the cluster wind mechanical 
energy output and affects particularly the cluster central regions, leading 
to frequent thermal instabilities that diminish the pressure and inhibit the 
exit of the reinserted matter. Instead matter accumulates there and is 
expected to eventually lead to gravitational instabilities and to further 
stellar formation with the matter reinserted by former massive stars. The 
main outcome of the calculations is that the critical line is almost two 
orders of magnitude or more, depending on the assumed value of $V_{A\infty}$, 
lower than when  only gas radiative cooling is applied. And thus, many massive 
clusters are predicted to enter the bimodal regime.  
\end{abstract}


\section{Introduction}

Data from Spitzer and Herschel satellites have unveiled type II supernovae as major producers of dust. Notable cases are those of the supernova 1987A \citep[see][]{Moseleyetal1989,Matsuuraetal2011}, the crab Nebula \citep{Gomezetal2012},  Cas A \citep{Hinesetal2004} as well as the undecided type Kepler supernova \citep[see][]{Reynoldsetal2007,Morganetal2003}, leading in all cases to dust masses, $M_d$, of the order of several tenths to a few solar masses. The  implication is that the condensation of refractory elements into dust is very efficient in the ejecta of core-collapse supernovae. This result is central to explain the large amount of dust ($\geq$ 10$^8$ M$_{\odot}$) present in high redshift ($z$ $\geq$ 6) galaxies \citep{Bertoldietal2003,Hinesetal2006} as any other significant dust producer, such as the winds from low-mass evolved stars,  requires of an evolution time comparable to the age of the Universe at that time 
\citep{MorganEdmunds2003}. The production of dust during the explosion of type II SN has also been investigated theoretically.
\citet{TodiniFerrara2001} and \citet{BianchiSchneider2007} found that the collapse of stars with masses in the range 12 to 40 M$_{\odot}$ with primordial metallicity, leads to the condensation of 0.08 M$_{\odot}$ $\leq$ M$_d$ $\leq$ 0.3 M$_{\odot}$ of dust per supernova (SN). These values increase by a factor of about three if  the metallicity is enhanced to solar values. \citet{Nozawaetal2003} considering SNe with masses up to 120 M$\odot$, estimated the ratio of dust mass to progenitor mass to be of the order of 0.02 - 0.05 and thus the calculated resultant dust masses are in 
excellent agreement with the values observed in young SN remnants (1987A, Cas  A, the crab nebula and Kepler). 

Here we show that the production of large quantities of dust in type II SN plays also a major role in the hydrodynamics of the matter reinserted in young and massive ($M_{SC}$ $\geq$ 10$^6$ M$_{\odot}$) superstar clusters (SSCs). In such clusters, given any reasonable IMF, one expects several tens of thousands of type II supernovae spread over the first 40 Myr of evolution. The main impact 
of this is on the cooling at infra-red frequencies expected from the dust immersed in the hot ($\sim 10^7$ K) thermalized ejecta, which in the adiabatic case eventually flows supersonically out of the cluster as a stationary cluster wind. The dust cooling of a hot gas was first envisaged by \citet{OstrikerSilk1973} who showed that the radiation from dust particles becomes larger than that generated by the gas in ionization equilibrium, including all of its possible radiative processes, at temperatures $\sim$ 10$^6$ K. The cooling law (erg cm$^3$ s$^{-1}$) then increases more than two orders of magnitude reaching a factor of 400 above the gas cooling when this is at  its minimum  ($\sim 10^7$ K). And from then onwards, going to even larger temperatures, up to 10$^9$ K,  it remains about two orders of magnitude above the bremsstrahlung cooling.    

Very similar results have been found by different authors for a variety of different conditions (size of grains, erosion, shock velocities, etc.) and applications (intergalactic matter in galaxy clusters, Seyfert galaxies, supernova and their remnants). Thorough studies are those of  \citet{BurkeSilk1974}; \citet{Draine1981}; \citet{DwekWerner1981}; \citet{Dwek1981,Dwek1987}; \citet{Smithetal1996}; \citet{MontierGiard2004} and more recently \citet{EverettChurchwell2010}; and \citet{Guillardetal2009}. In all of them, as in the original \citet{OstrikerSilk1973}, dust is by far the main coolant at high temperatures. Here gas and dust cooling are added and applied to the thermalized matter reinserted by the population of massive stars in  coeval young super star clusters. 

Section 2 shows how the expected dust sputtering time and the SN rate in massive clusters, confabulate to lead to a steady state in which as dust is eroded within a cluster, it is replenished back by further supernovae. This also leads to a dust to gas mass ratio $\sim 10^{-3} - 10^{-2}$. Section 3 deals with the relevance of strong radiative cooling on the matter reinserted and thermalized during the early evolutionary stages of coeval super star clusters.  Here we  show how to find the location of the threshold line in the mechanical luminosity or cluster mass {\it vs} size of the cluster parameter space, when dust radiative cooling is taken into consideration. Such a critical line separates 
stationary cluster winds from the bimodal cases in which strong cooling 
within the cluster volume leads to mass accumulation and to further star 
formation.  This was proposed by \citet{GTTetal2007} based on 1D numerical 
simulations and confirmed with 2D results by \citet{Wunschetal2008}. Our main 
conclusions are given in section 4. 

\section{The stationary presence of dust within young stellar clusters} 

As pointed out by \citet{Smithetal1996} the inclusion  of dust cooling into a time dependent  hydrodynamic evolution requires to  
closely follow how dust is eroded and how this changes the gas phase abundances. The bookkeeping has to care about the different dust constituents, the time dependent dust size distribution, and how the elemental abundances are enhanced as dust is eroded. Erosion or the sputtering of dust grains is caused mainly by the bombardment of energetic particles such as protons, helium nuclei and other dust grains. Electrons are usually not considered due to their low efficiency \citep{DwekArendt1992}.  Here we consider the stationary situation that results from frequent SN explosions  within massive SSCs.  The time evolution outside the stellar cluster, within the cluster wind and its interaction with the ISM  requires surely the time dependent tracking of dust sputtering and the variations of gas phase abundances. However, within the cluster volume the average dust mass production and dust mass depletion rates must be in balance. Thus, one simply must account for the dust production rate $\dot M_d$ as well 
as for the rate at which all other processes may lead to its depletion within the flow. 
Among these, the most obvious are: sputtering and, if the dust is coupled to 
the gas, its exit as a constituent of the cluster wind. Such considerations lead to the relation:
\begin{equation}
      \label{eq1}
{\dot M}_d - \frac{M_d}{\tau_d} - {\dot M}_g \frac{M_d}{M_g} = 0 ,
\end{equation}
where $\tau_d$ is the characteristic dust destruction time scale. The various terms in equation (\ref{eq1}) represent  the dust mass production rate, the dust destruction rate and the dust mass loss rate due to the outflow of the reinserted matter as a stationary cluster wind. Hereafter we will assume that the dust mass input rate ${\dot M}_d$ is in direct proportion to the gas mass input rate, 
${\dot M}_d = \alpha {\dot M}_g$, as dust is here assumed to be injected (via SN) together with the gas (which comes from winds and SN). 
One can make use of equation (\ref{eq1}) to obtain the expected dust to gas mass ratio
within the considered cluster volume:
\begin{equation}
      \label{eq2}
Z_d = \frac{M_d}{M_g} = \frac{\alpha \tau_d {\dot M}_g}{M_g} 
                  \left(1 + \frac{\tau_d {\dot M}_g}{M_g}\right)^{-1} ,
\end{equation}
The mass of the 
reinserted gas within the star cluster radius ($R_{SC}$) is:
\begin{equation}
      \label{eq3}
M_g = 4 \pi \int_0^{R_{SC}} \rho(r) r^2 {\rm d} r =
      \frac{f_g R_{SC} {\dot M}_g}{3 c_s} . 
\end{equation}
The factor $f_g \approx 2$ in equation (\ref{eq3}) takes into account the fact that the gas density within the cluster is not uniform, it drops from the center outwards to reach the value  $\rho_s = {\dot M}_{SC} / 4 \pi c_s R^2_{SC}$ at the star cluster surface, where $c_s$ is the sound speed at the star cluster edge ($c_s \approx v_{\infty} / 2$), $v_{\infty}$ is the wind terminal speed \citep{Cantoetal2000, Palousetal2013}, ${\dot M}_{SC}$ is the mass deposition rate within the cluster volume. Note that if  the stellar distribution is not 
homogeneous, as here assumed, the gas density of the reinserted matter falls even more steeply \citep[see][]{Silichetal2011, Palousetal2013} making the dust sputtering time even longer.  Combining equations (\ref{eq2}) and (\ref{eq3}), one can obtain:
\begin{equation}
      \label{eq4}
Z_d = \frac{M_d}{M_g} = \frac{3 \alpha \tau_d c_s}{f_g R_{SC}} 
                  \left(1 + \frac{3 \tau_d c_s }{f_g R_{SC}}\right)^{-1} .
\end{equation}
The dust  life-time against sputtering at temperatures above 10$^6$ K is given by \citet{DraineSalpeter1979}: $\tau_{d} = 10^6 a / n (yr) =
B / n (sec)$, where $B = 3.156 \times 10^{13} a$, $a$ is the size of the considered dust grains in $\mu m$ and $n$ is the average nucleon number density in the gaseous phase. Substituting this relation into equation (\ref{eq4}) and taking into account that the average density of the reinserted matter is within a factor $f_g$ larger than at the star cluster surface ($n = f_g {\dot M}_{SC} / 2 \pi \mu_i v_{\infty} R^2_{SC}$), one can obtain: 
\begin{equation}
      \label{eq6}
Z_d = \frac{M_d}{M_g} = \frac{3 \pi \alpha B \mu_i R_{SC} v^4_{A\infty}}
                       {2 f^2_g L_{SC}}  \frac{L_{out}}{L_{SC}}
      \left(1 + \frac{3 \pi B \mu_i R_{SC} v^4_{A\infty}}
      {2 f^2_g L_{SC}} \frac{L_{out}}{L_{SC}}\right)^{-1} ,
\end{equation}
where the ratio of the energy flux through the star cluster surface to the star cluster mechanical luminosity, $L_{out} / L_{SC} = v^2_{\infty} /v^2_{A\infty}$, $v_{A\infty} = 2 L_{SC} / {\dot M}_{SC}$ is the adiabatic wind terminal speed, $\mu_i = 14 m_H / 11$ is the mean mass per nucleon and $m_H$ is the proton mass. In the non-radiative wind model $L_{out}$ is equal to the star cluster mechanical luminosity and $L_{out}/L_{SC} = 1$. However in radiative winds $L_{out} < L_{SC}$ and $v_{\infty} < v_{A\infty}$ due to radiative losses of energy inside the star cluster volume.

Equation (\ref{eq6}) shows that the dust to gas mass ratio $Z_d$ must be smaller in more compact and more energetic clusters as in such clusters the average density of the reinserted matter is larger and thus the dust sputtering time is smaller. The average gas number density also increases in clusters with smaller $v_{A\infty}$ which leads to a strong dependence of $Z_d$ on the adiabatic wind terminal speed. Note that the value of $Z_d$ can never exceed that in the injected matter and it sets the upper limit for the dust over gas mass ratio inside the cluster: $Z_d < \alpha$. Figure 1 presents the expected dust to gas mass ratio in clusters with different sizes, 
mechanical luminosities and adiabatic wind terminal speeds containing dust grains of different sizes. Taking a conservative point of view, it was assumed that stellar winds and supernovae contribute similar amounts of gas \citet{Leithereretal1999} and thus the dust to gas mass ratio in the injected matter is two times smaller than the lowest value obtained by \citet{Nozawaetal2003}
for supernovae 
ejecta: $\alpha = 0.01$. $L_{out} / L_{SC}$ was set to values consistent with 
those found at the critical line that separates stationary and thermally 
unstable solutions: 0.69 for $v_{A\infty} = 1000$~km s$^{-1}$, 0.77 for 
$v_{A\infty} = 1500$~km s$^{-1}$ and 0.82 for $v_{A\infty} = 
2000$~km s$^{-1}$, respectively.
\begin{figure}[htbp]
\vspace{15.0cm}
\includegraphics{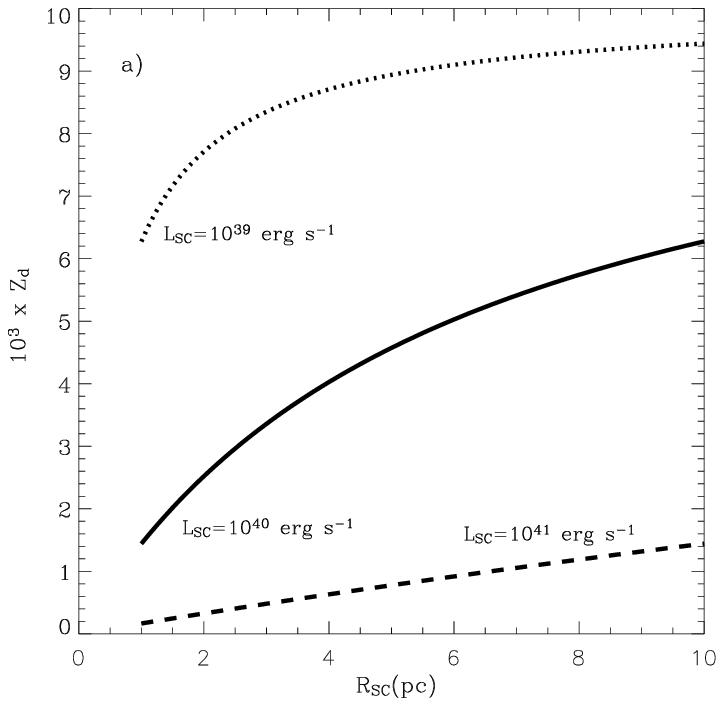}
\includegraphics{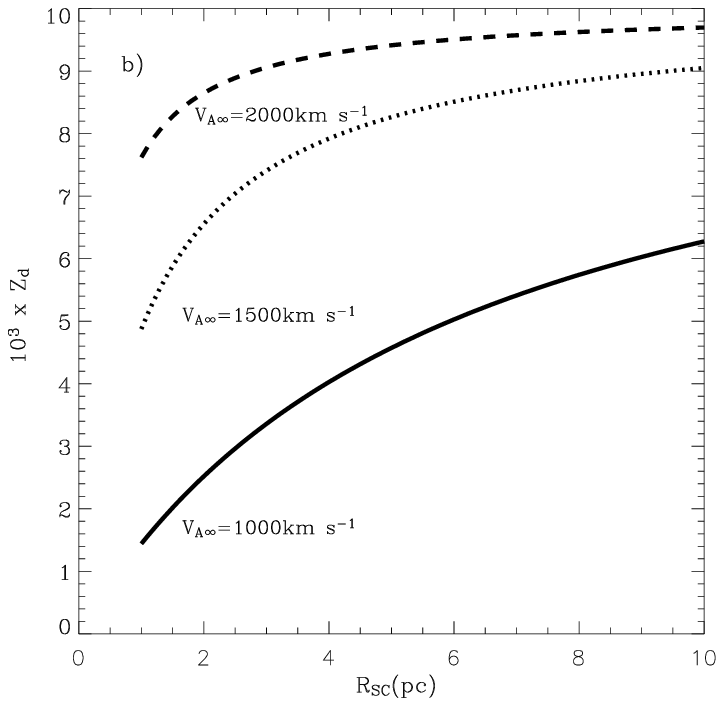}
\includegraphics{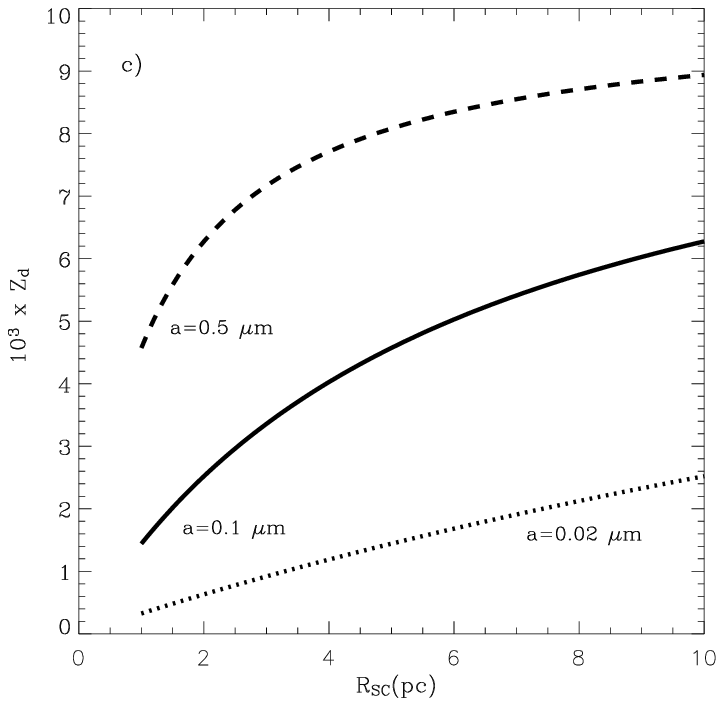}
\caption{The expected dust to gas mass ratio inside young stellar clusters. 
Panel a presents the dust to gas mass ratio, $Z_d$, calculated by means of 
equation (\ref{eq6}) for clusters with different mechanical luminosities 
($L_{SC} = 10^{40}$~erg s$^{-1}$, $L_{SC} = 10^{41}$~erg s$^{-1}$ and 
$L_{SC} = 10^{39}$~erg s$^{-1}$ - solid, dashed and dotted lines, 
respectively) as a function of the star cluster radius, when the assumed size 
of the dust grains is $a = 0.1 \mu$m and the adiabatic wind terminal 
speed is $V_{A\infty} = 1000$~km s$^{-1}$. Panel b displays the same ratio 
for clusters with a mechanical luminosity $L_{SC} = 10^{40}$~erg s$^{-1}$ and 
different adiabatic wind terminal speeds: $v_{A\infty} = 1000$~km s$^{-1}$,  
$v_{A\infty} = 1500$~km s$^{-1}$ and  $v_{A\infty} = 2000$~km s$^{-1}$, 
solid, dotted and dashed lines, respectively. Panel c shows $Z_d$ in cluster 
with $L_{SC} = 10^{40}$~erg s$^{-1}$ and $v_{A\infty} = 1000$~km s$^{-1}$ when
dust grains have different radii: $a = 0.1 \mu$m, $a = 0.5 \mu$m and 
$a = 0.02 \mu$m: solid, dashed and dotted lines, respectively.}
\label{fig1}
\end{figure}

Note that only in the most  energetic and compact clusters here considered the value of $Z_d$ falls below $10^{-3}$ (dashed line in panel a) while in other clusters, either with different mechanical luminosities (panel a) or different wind terminal speeds (panel b), the dust to gas mass ratio falls always in the range $10^{-3} \le Z_d \le 10^{-2}$ unless the size of the dust grains is very small (panel c). The implication is that within a star cluster volume, as dust grains are sputtered they are replenished back by further supernovae, causing a stationary presence of dust. Such a condition points at the importance of dust in the thermodynamics of the matter reinserted within SSCs. 
 
\subsection{The dust cooling law}

Following \citet{DwekWerner1981}, we have calculated the cooling function due to gas-grain collisions for a plasma with a normal chemical composition (one He atom per every ten H atoms) as:
\begin{eqnarray}
\Lambda_d = \frac{n_d}{n_en_H} H_{coll} = 
\frac{1.4 m_{H} Z_d}{\langle m_d \rangle} \left(\frac{32}{\pi m_{e}}\right)^{1/2} \pi a^2(kT)^{3/2}  
\left[h_{e} + \frac{11}{23} \left( \frac{m_{e}}{m_{H}} \right)^{1/2}h_{n}   \right] ,
\end{eqnarray}
where $n_d$, $n_e$ and $n_H$ are the dust, electron and hydrogen particle number density, $H_{coll}$ is the heating rate of a single grain due to collisions of incident gas particles. Functions $h_{e}$ and $h_{H}$ are the effective grain heating efficiencies due to incident electrons and nuclei, respectively:
\begin{eqnarray}
h_{e}=1-\int^{\infty}_0 (z+x_{e})[(z+x_{e})^{3/2}-x_{e}^{3/2}]^{2/3} \mbox{d}z ,
\end{eqnarray}
\begin{eqnarray}
h_{n}= 1-(1+E_{H}/2kT) exp(-E_{H}/kT)  , 
\end{eqnarray}
where $m_{H}$ is the proton mass, $\langle m_d \rangle=4/3\pi \rho_{d} a^3 $ is the mass of the dust grains, $x_{e}=E_{e}/kT$, $E_{e}=23 a^{2/3}$($\mu$m) keV and $E_{H}=133 a$($\mu$m) keV. We neglected the dust grains charge \citep{Smithetal1996} and assumed that all of them have the same size $a$ and density $\rho_{d}= 3$ g cm$^{-3}$. 

Figure 2 shows the cooling curve obtained for dust grains with $a=0.1$ $\mu$m and different $Z_{d}$, compared to the gas cooling function. The curve for $Z_{d}=6\times 10^{-3}$ reproduces that of \citet[their Figure 2]{DwekWerner1981}.  
\begin{figure}[htbp]
\plotone{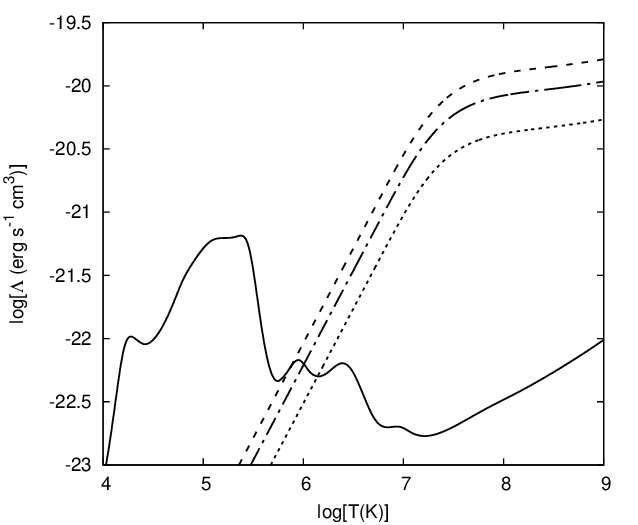}
\caption{The cooling function. The dust cooling law as a function of 
temperature for different values of $Z_d$: 3$\times 10^{-3}$, 6$ \times 
10^{-3}$ and 9$\times 10^{-3}$ (dotted, dash-dotted and dash lines, 
respectively) compared to the interstellar cooling law (solid line).}  
\label{fig2}
\end{figure}

\section{Strong radiative cooling and the location of the threshold line.}

As shown in a recent series of papers \citep{Silichetal2004, GTTetal2005a, GTTetal2007, Wunschetal2008, Wunschetal2011}, the effects of gas radiative cooling on the hydrodynamics of the matter reinserted by winds and supernova explosions within homogeneous, young, massive and compact stellar clusters may lead to a bimodal solution, instead of the powerful stationary cluster winds expected in adiabatic calculations \citep[see][]{ChevalierClegg1985}. 

In the latter, all the deposited matter ($\dot M_{SC}$) ends up streaming supersonically into the ISM as a stationary cluster wind. This is a situation in which the amount of deposited matter, $\dot M_{SC}$, leaves the cluster as it is reinserted by massive stars: $\dot M_{SC} = 4 \pi R_{SC}^2 \rho_{s} c_{s}$.  For this to happen, the flow has to present several special features such as: the stagnation point ($R_{st}$; the place where the flow velocity equals zero km s$^{-1}$) has to be at the center of the cluster. The reinserted matter is then steadily accelerated outwards in response to  pressure gradient, to reach the local sound speed ($c_{s}$) right at the cluster surface ($R_{SC}$). In such a case,  the large pressure difference with the surroundings  accelerates the flow even further for it to reach supersonic terminal speeds and conform the cluster wind. In the bimodal regime (cases above the threshold line), the main difference is that  strong radiative cooling becomes important particularly within 
the cluster densest central regions and this causes zones within the flow to frequently become thermally unstable. 
The instabilities cause a sudden loss of pressure and this  inhibits the outward acceleration required to form part of the cluster wind. Instead, the unstable parcels of gas begin to accumulate while condensing under the action of the surrounding hot matter, in search of a rapid re-establishment of an even pressure. That leads eventually to a gravitational instability and thus to an extreme  positive star formation feedback condition, in which further generations of stars form within the cluster volume with the matter reinserted
by  massive stars \citep{GTTetal2005b, GTTetal2005a}.    

As we have shown \citep{GTTetal2007, Wunschetal2008}, only clusters located above the threshold line, in the mechanical luminosity (or cluster mass) {\it vs} size diagram, are able to undergo the bimodal solution. In all such clusters with a homogeneous stellar density distribution, the frequent and recurrent thermal instabilities promote the exit of the stagnation radius out of the cluster center and make it approach the cluster boundary and the more so, the further above the threshold line the considered clusters are. Matter 
reinserted in the volume between the stagnation radius and the cluster edge still manages to accelerate and reach the sonic point right at the cluster edge and thus composes a wind.  

Here we search for the location of the threshold line when cooling by dust is also included following the procedure given by 
\citet{GTTetal2007}. The integration of the hydrodynamic equations for the flow demands knowledge of the temperature at the stagnation point, $T_0$, which may be found by iteration until the sonic point is accommodated at the cluster edge. 
 
In this way, a unique stagnation temperature, from the branch of possible temperatures, is selected  and once the selected temperature corresponds to the maximum pressure, the critical energy has been reached. We can thus select the cluster parameters ($R_{SC}$, $V_{A\infty}$, $L_{SC}$) as well as the dust parameters ($a$ and $\rho_{d}$) and through equation (5) find $Z_d$. This allows one to calculate the dust cooling curve and with this and by iteration, find the largest mechanical energy for which a stationary wind results with its stagnation point at the cluster center and that reaches its local sound speed at the cluster surface.  

Figure 3 shows the location of the threshold line derived when using only gas cooling from a gas in collisional ionization equilibrium 
\citep[see][]{Silichetal2004, GTTetal2007} compared to the critical line when one adds gas and dust cooling. Dust cooling lowers the threshold cluster mechanical luminosity (or cluster mass) by about 2 dex or more, depending on the assumed value of $V_{A\infty}$. Thus many massive ($M_{SC} \geq 10^5$ M $_\odot $) clusters which appear as quasi-adiabatic when one considered only gas cooling,
are now well above the threshold line, and thus in the bimodal regime. The calculations show that at the critical line the fraction of the injected energy that clusters return to the interstellar medium, $L_{out}/L_{SC} \approx 0.69$, and as was shown by \citet{GTTetal2007}, this decreases monotonically as one selects more massive clusters with a larger excess energy above the critical value ($L_{SC} / L_{crit}$), \citet[see Figure 5 of][]{GTTetal2007}. Thus, radiative cooling may strongly deplete the mechanical energy output from massive clusters and thus reduce their negative star formation feedback into the ISM.  
\begin{figure}[htbp]
\plottwo{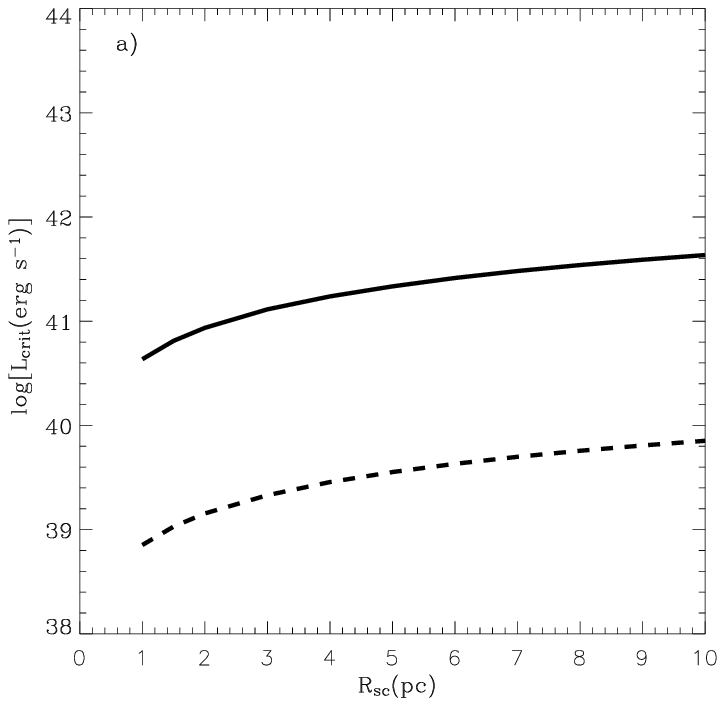}{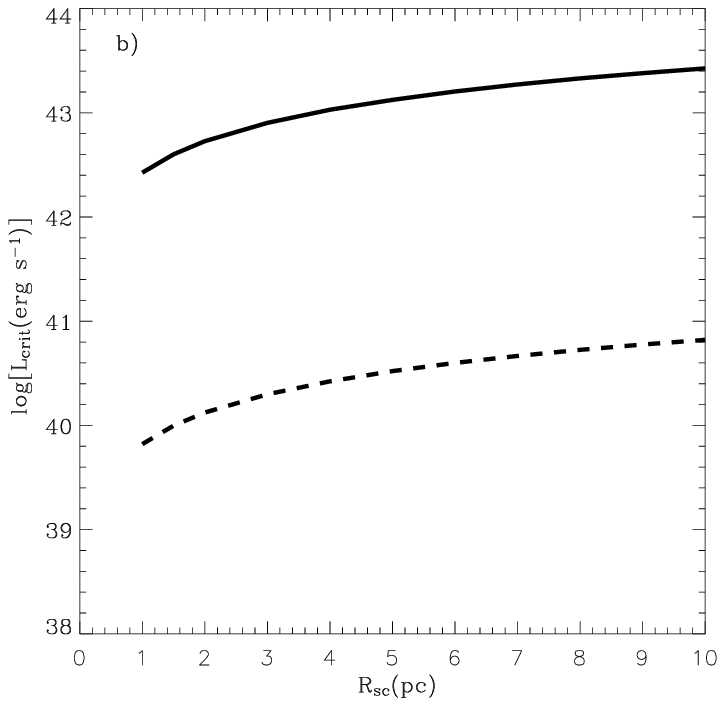}
\caption{The threshold mechanical luminosity. Panels a and b present 
$L_{crit}$ calculated for $V_{A\infty} = 1000$ and $2000$~km s$^{-1}$ under 
the assumption of pure gas  cooling (solid lines) and when gas and dust  
cooling are added (dashed lines). In all cases dust grains were assumed to 
have a radius  $a = 0.1 \mu$m.}
\label{fig3}
\end{figure}

Figure 4 displays the run of velocity, as well as temperature and density, 
for  clusters with 1 pc and 10 pc radius and a luminosity $L_{SC} = 7.13 
\times 10^{38}$ erg s$^{-1}$ and $L_{SC} = 7.13 \times 10^{39}$ erg 
s$^{-1}$, respectively. Such values  places them at the critical line 
when dust and gas cooling are considered. The figure compares the run of all 
of these variables with those derived for the quasi-adiabatic solution when 
only gas radiative cooling is considered. The large central temperature 
depressions  at the critical line and the slow acceleration of the reinserted 
matter within the cluster are indicative of the last stationary solution 
found before crossing the critical line.

The presence of dust in the thermalized plasma within young SSCs provides a natural explanation to the observational and theoretical studies that have derived a strongly  reduced mechanical energy output from some of the massive clusters in the galaxy M82  in order to account for  the size of the compact HII regions that surround them \citep[see][]{Smithetal2006, Silichetal2007, Silichetal2009}. 
Further evidence for the bimodal solution and the positive star formation feedback may come from the interpretation of the abundance variations from star to star in massive globular clusters (GC), such as $\omega$ Cen in the Milky Way where different generations of stars are enriched by different contributors and among these by core collapse SNe \citet{Grattonetal2004} or from observations of the most massive GCs in M31 where the run of $\alpha$ to Fe-peak elements is consistent with a primordial enrichment from stars with masses larger than 10 M$_\odot$ \citep{Meylanetal2001}. The super-solar $\alpha$ to Fe element ratios indicating on the fast gas enrichment by Type II supernovae were also detected by \citet{Larsenetal2006} who have obtained the near-IR H- and K-band spectra and provided a detailed abundance analysis of a young massive cluster in the nearby spiral galaxy NGC 6946. The presence of the hot dust component ($\sim$ 800 K) is required in order to fit the UV to near IR spectral energy 
distribution of SSCs 1 and 2 in the low metallicity galaxy ($Z = 1/40 Z_\odot$) SBS 0335-052 \citep[see Figure 8 of]{Reinesetal2008}.
\begin{figure}[htbp]
\plotone{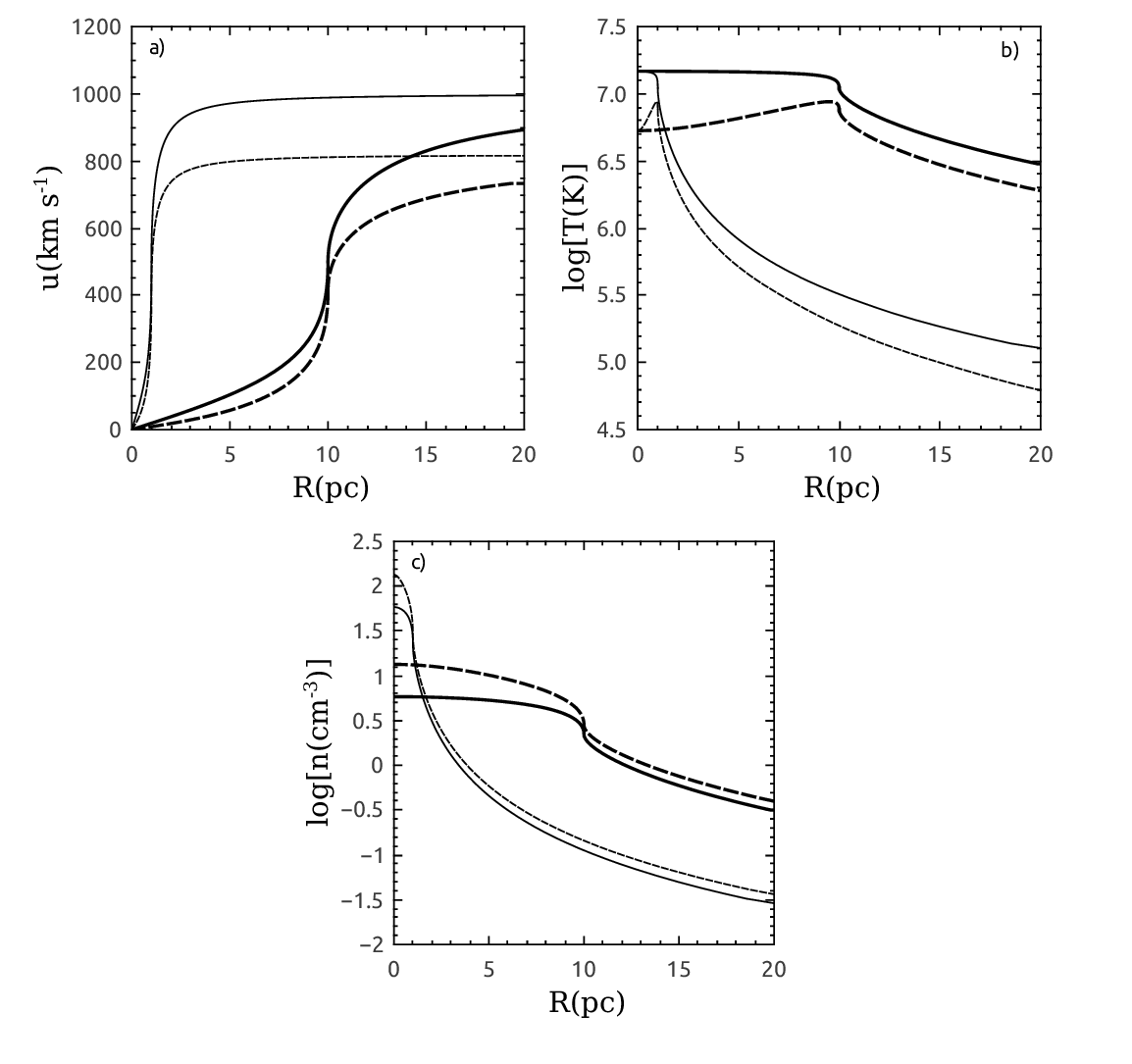}
\caption{The impact of dust radiative cooling on the distribution of the 
hydrodynamical variables. Panels a, b and c present the wind velocity, 
temperature and density, respectively. The thin and thick lines display the 
results of the calculations for clusters with $R_{SC} = 1$~pc and $R_{SC} = 
10$~pc. The dashed and solid lines correspond to models with and
without dust cooling, respectively. The same adiabatic wind terminal speed
and grain sizes were used in all calculations: $V_{A\infty} = 1000$~km 
s$^{-1}$ and $a = 0.1 \mu$m.}  
\label{fig4}
\end{figure}

\section {Conclusions}

Observational and theoretical evidences point at an efficient condensation of refractory elements into dust in the ejecta of core-collapsed SN. The several tens of thousands of type II SN expected in young and massive SSCs has led us to postulate a continuous presence of dust in the SSCs volume and conclude that the likely range of dust to gas mass ratios is $Z_d \sim 10^{-3} - 10^{-2}$, which has allowed us to calculate the dust cooling law within young and massive SSCs. Dust cooling effectively lowers the location of the 
critical line which separates clusters with stationary outflows from those evolving in the bimodal regime. The new location implies that young, massive and compact SSCs with masses $M_{SC} > 10^5$ M$_\odot$ experience the bimodal solution.         

The implication of assuming gas and dust cooling is thus that for young, massive and compact stellar clusters only a fraction of their mechanical luminosity, as inferred from synthesis models (as in SB99), impacts the surrounding ISM. Massive, compact and young clusters thus inject into the ISM only a fraction of their reinserted matter and of their processed metals and  with a velocity which is more strongly diminished the further above the critical line the considered clusters are. Meanwhile, the matter reinserted within the thermally unstable volume is continuously reprocessed into dense compact clouds compressed by the surrounding hot plasma and photoionized by the parental cluster Lyman continuum. Mass accumulation would lead to surpass the Jeans instability limit and  to an extreme mode of positive star formation feedback within the cluster volume, where the matter reinserted by massive stars leads to new stellar generations.

\acknowledgments

We thank our anonymous referee for multiple suggestions which led to the
clarification of our results. This study has been supported by CONACYT - 
M\'exico, research grants 167169 and 131913, by the bilateral research
project "Violent star formation" between CONACYT - M\'exico and the Academy 
of Sciences of the Czech Republic and by the Spanish Ministry of Science and 
Innovation under the collaboration ESTALLIDOS (grants AYA2007-67965-C03-01 and 
AYA2010-21887-C04-04). JP and RW also acknowledge support from the projects
RVO 67985815 of the Academy of Sciences of the Czech Republic and P209/12/1795
of the Czech Science Foundation.

\bibliographystyle{apj2}
\bibliography{dusty}


\end{document}